\ifdefined\label
\let\manuscriptKernelLabel\label
\fi
\documentclass[aps,prx,reprint,superscriptaddress,floatfix]{revtex4-2}

\usepackage{graphicx}
\graphicspath{{images/}}
\usepackage{amsmath,amssymb,amsfonts}
\usepackage{bm} 
\usepackage{color}
\usepackage{xcolor}
\usepackage{soul}
\setstcolor{black}
\usepackage{booktabs}
\usepackage[utf8]{inputenc}
\usepackage{comment}

\usepackage{multirow}
\usepackage{appendix}
\usepackage{ragged2e}
\usepackage{physics}

\usepackage{enumitem}
\setlist{leftmargin=3.5mm}

\usepackage{placeins}
\usepackage[normalem]{ulem}
\usepackage{tikz}
\usetikzlibrary{decorations.markings}
\usetikzlibrary{calc}
\usetikzlibrary{arrows.meta, positioning}
\ifdefined\manuscriptKernelLabel
\let\label\manuscriptKernelLabel
\makeatletter
\let\ltx@label\manuscriptKernelLabel
\makeatother
\fi
\usepackage{hyperref} 
\newcommand{\ay}[1]{#1}

\newcommand{\tabletext}{\normalfont\rmfamily\linespread{1}\fontsize{9}{11}\selectfont}
\begin{document}
\color{black}
\raggedbottom

\title{Post-Optimization Subspace Corrections to Quantum Ansätze} 
\author{Prince Frederick Kwao} 
\affiliation{Department of Chemistry, University of North Dakota, {Grand Forks}, ND 58202, USA}
\author{Bharath Sambasivam}
\affiliation{Department of Physics, Virginia Tech, Blacksburg, VA 24061, USA}
\affiliation{Virginia Tech Center for Quantum Information Science and Engineering, Blacksburg, VA 24061, USA}
\author{Jason Saroni}
\affiliation{Department of Physics, Virginia Tech, Blacksburg, VA 24061, USA}
\affiliation{Virginia Tech Center for Quantum Information Science and Engineering, Blacksburg, VA 24061, USA}
\author{Srivathsan Poyyapakkam Sundar}
\affiliation{Department of Chemistry, University of North Dakota, {Grand Forks}, ND 58202, USA}
\author{Edwin Barnes}
\affiliation{Department of Physics, Virginia Tech, Blacksburg, VA 24061, USA}
\affiliation{Virginia Tech Center for Quantum Information Science and Engineering, Blacksburg, VA 24061, USA}
\author{Ayush Asthana}
\email{ayush.asthana@und.edu}
\affiliation{Department of Chemistry, University of North Dakota, {Grand Forks}, ND 58202, USA}

\begin{abstract}
\ay{Circuit} depth and measurement count are the main limitations for chemical applications of the leading variational quantum algorithms on quantum computers. {We show that both resource requirements can be reduced by augmenting adaptive variational quantum algorithms with post-optimization corrective techniques. We focus on two such techniques---Hamiltonian-moment Krylov subspace diagonalization and the identity-augmented quantum self-consistent equation of motion (q-sc-EOM)---applied to optimized ADAPT-VQE states without further ansatz growth or nonlinear optimization.} Across molecular systems studied with up to 20 qubits, first-order Krylov correction reduces the optimized iterations (and the corresponding number of adaptive layers) needed to reach the $1.6~\mathrm{m}E_h$ threshold by up to approximately sixfold, with a general trend toward larger reductions at higher determinant R\'{e}nyi entropy. {We also find measurement-cost reductions of up to approximately fortyfold using a suitable resource-cost model.} Identity-augmented q-sc-EOM improves molecular potential-energy curves obtained from {truncated ADAPT-VQE ans\"atze with limited expressivity} and recovers accuracy from {states produced by sampled operator ranking} without additional gradient-based operator selection, {providing} an advantage when small pool gradients make selection measurement intensive. Together, these post-optimization corrections offer routes to improve energy accuracy and reduce adaptive circuit growth in quantum algorithms as quantum hardware advances.
\end{abstract}
\maketitle

\sloppy

\section{Introduction}

{Quantum computing offers an attractive route to solving challenging molecular electronic-structure problems~\cite{mcardle2020quantum,cao2019quantum,aspuru2005simulated,liu2022prospects,stair2021simulating,smart2021quantum,Smart2022AcceleratedCQE}.} {Representative challenges include multiple-bond dissociation, transition-metal chemistry, photochemical processes, and the characterization of potential-energy surfaces near conical intersections, for which accurate electronic-state energies are essential~\cite{cao2019quantum,kohn2007can,thomas2021complex,kjonstad2017crossing,tajti2018accuracy,wang2024quantum}.} {For near-term quantum hardware, variational quantum algorithms offer a way to balance resource requirements and energy accuracy through the choice and optimization of the ansatz~\cite{cerezo2021variational,Preskill2018}.}
Peruzzo \emph{et al.} introduced the variational quantum eigensolver (VQE) for quantum-chemistry applications~\cite{peruzzo2014variational}. VQE may use a fixed ansatz, such as unitary coupled cluster with singles and doubles (UCCSD), whose operator content is predetermined~\cite{BARTLETT1989133,romero2018strategies}. Alternatively, the adaptive derivative-assembled problem-tailored VQE (ADAPT-VQE) algorithm constructs a problem-specific ansatz iteratively{~\cite{grimsley2019adaptive,grimsley2025challenging,Saroni2026TEPID}}. At each outer iteration, it evaluates operator-pool gradients, appends the operator or compatible operator set selected by the prescribed rule, and reoptimizes all accumulated parameters. This adaptive construction can avoid operators that contribute little to local energy decrease and can produce shallower, problem-tailored circuits~\cite{tang2021qubit,Shkolnikov2023SymmetryRoadblocks}. ADAPT-VQE also avoids optimization issues such as local traps and barren plateaus that plague some fixed-ansatz VQEs~\cite{Grimsley2023ADAPTLandscapes}.

However, the practical cost of both VQE and ADAPT-VQE is dominated by repeated estimation of expectation values. Molecular Hamiltonians, operator-pool gradients, and parameter reoptimization can require many measurements with sufficient precision~\cite{Huggins2021,crawford2021efficient,Anastasiou2023Gradients,Gonthier2022MeasurementRoadblock,Nykanen2025MeasurementADAPT}. Moreover, recent molecular benchmarks reported an approximately exponential increase in adaptive iteration requirements with determinant R\'{e}nyi entropy~\cite{hagelueken2026exponentialscalingbarriersvariational}. At equilibrium geometries, shallow ansätze may approximate the ground state accurately, but their performance often degrades as bonds are stretched and static correlation increases. Near-degenerate determinants then produce a multireference character that may not be adequately represented by a shallow variational manifold~\cite{lyakh2012multireference,jctc.9b01125}. Extending the circuit can recover additional correlation but increases optimization and implementation costs{~{\cite{meitei2021gate,asthana2022minimizing,grimsley2019adaptive,MagoulasEvangelista2023,Patra2024ASC}}}; adaptive growth can also exhibit apparent energy saturation when the remaining pool gradients become small~\cite{Bertels2022SymmetryBreakingADAPT,Stadelmann2025GradientTroughs}.

Subspace diagonalization offers an alternative to further variational optimization: the optimized state provides a reference for an enlarged linear subspace, allowing additional correlation to be recovered while keeping the circuit parameters fixed~{\cite{motta2024subspace,Takeshita2020,mcclean2020decoding,mcclean2017hybrid,urbanek2020chemistry,Hirsbrunner2024CSVQE,Boyd2025ShadowsQSE}}. However, additional measurements and poorly conditioned overlap matrices can limit the practical benefit of subspace diagonalization~\cite{mcclean2017hybrid,cortes2022quantum,epperly2022theory,Zhang2024MeasurementEfficientKrylov,Kirby2024KrylovErrors,Nakamura2024AdaptiveMeasurement}.

Hamiltonian-moment corrections have been developed for error mitigation through the Lanczos scheme of Suchsland et al.~\cite{Suchsland2021Lanczos} and the classical-shadow implementation of Fischer et al.~\cite{Fischer2025ShadowsQSE}. Related ground-state energy approaches include the quantum connected-moment work of Kowalski and Peng~\cite{kowalski2020quantum}. \textcolor{black}{Claudino et al. combined shallow ADAPT-VQE references with low-order connected-moment and Peeters--Devreese--Soldatov (PDS) energy estimates, demonstrating improvements over continued ADAPT optimization in selected examples and reducing measurement requirements through term reuse and screening~\cite{Claudino2021ConnectedMoments}.} \textcolor{black}{Nevertheless, the broader potential of subspace diagonalization to support adaptive variational algorithms warrants systematic investigation. This requires understanding how reductions in adaptive iterations and parameter optimization balance the additional measurement requirements, how these benefits scale with problem complexity, and how they depend on the subspace construction. Beyond improving energy accuracy, such approaches may also alleviate limitations associated with continued ansatz growth and small operator-pool gradients.}

\textcolor{black}{In this study, we assess the improvements offered by two promising subspace diagonalization approaches, Krylov and identity-augmented quantum self-consistent equation-of-motion (q-sc-EOM), in terms of reductions in adaptive iterations, measurement requirements, and performance under finite-shot operator ranking.} For the Krylov study, we examine successive optimized states obtained with a coupled exchange operator (CEO) pool~\cite{Ramoa2025CEO} and TETRIS selection~\cite{PhysRevResearch.6.013254} across molecular systems containing up to 20 qubits. \textcolor{black}{For the first-order (K$_1$) and second-order (K$_2$) Krylov corrections, we evaluate expectation values of Hamiltonian powers with respect to the optimized ADAPT-VQE state to generate Hamiltonian moments. These moments, required through $\langle\hat{H}^{3}\rangle$ for K$_1$ and $\langle\hat{H}^{5}\rangle$ for K$_2$, are then used to construct the projected Hamiltonian and overlap matrices for subspace diagonalization.} \textcolor{black}{We quantify the adaptive iterations required to reach $1.6~\mathrm{m}E_h$ accuracy.} Across the systems studied, K$_1$ reduces the adaptive iterations needed to reach the target accuracy by up to approximately sixfold, with larger benefits generally observed at higher determinant R\'{e}nyi entropy. \textcolor{black}{A resource-cost model used in this study gives reductions of up to approximately fortyfold in measurement cost.}

\textcolor{black}{As a second complementary strategy, building on the q-sc-EOM formulation~{\cite{asthana2023quantum,kumar2023quantum,sundar2026molecular,Windom2026Triples}}, we augment the singles-and-doubles excitation-operator manifold with the identity operator, allowing the optimized ADAPT-VQE reference to mix with the excitation-generated states through subspace diagonalization. This provides a corrected ground-state energy and wave function without further ansatz growth or parameter reoptimization, while retaining an analytically known identity overlap matrix. {We show that, for some systems, this construction can improve potential-energy curves obtained from truncated ADAPT-VQE references (representing ans\"atze with limited expressivity) and enable earlier attainment of the target accuracy for states generated under finite-shot operator ranking.} {We also show that the previously proposed Davidson algorithm and basis-rotation grouping (BRG) can reduce the modeled measurement requirements of identity-augmented q-sc-EOM ground-state corrections, extending their use in excited-state q-sc-EOM~\cite{kim2023two,Huggins2021,sundar2026molecular}.}}

This paper is structured as follows: Sec.~\ref{sec:theory} briefly introduces the theory of Krylov subspaces and q-sc-EOM. \textcolor{black}{Sec.~\ref{sec:computational_details} provides computational details relevant to the results.} Reductions in ADAPT-VQE iterations using Krylov subspaces and corrections to ground-state energies using q-sc-EOM are presented in Sec.~\ref{sec:results_and_discussion}. Sec.~\ref{sec:conclusion} contains concluding remarks.

\FloatBarrier
\section{Theory}
\label{sec:theory}
\subsection{Krylov subspaces}
Quantum Krylov subspaces are built by iteratively applying a function of the Hamiltonian to a reference state, producing the Krylov vector at iteration $i$~{\cite{Parrish2019QFD,jctc.9b01125}}:
\begin{align}
    \ket{\Psi_{i}} \propto \big(f(\hat{H})\big)^i\ket{\Psi_{0}},
\end{align}
where $|\Psi_0\rangle$ is a reference wave function, taken here to be the optimized ADAPT-VQE state at the iteration being corrected. 
Choosing $f(\hat H)=\hat H$ generates a Hamiltonian-power Krylov space, which underlies the Lanczos method after orthonormalization. {For a pure reference, exact Hamiltonian moments, and linearly independent Krylov vectors, the K$_1$ ground-state energy coincides with the two-dimensional Lanczos correction of Suchsland et al.~\cite{Suchsland2021Lanczos} and the PDS(2) estimate used by Claudino et al.~\cite{Claudino2021ConnectedMoments}.} Related imaginary-time and cumulant-based approaches use different constructions~\cite{GarnetNP2020,Vallury2020QCM}. Once the Krylov vectors are created, the Hamiltonian can be diagonalized within the subspace by solving the generalized eigenvalue equation,
\begin{align}
    \mathbf H\mathbf C=\mathbf S\mathbf C\mathbf E.
\end{align}
Here, $\mathbf C$ contains the eigenvectors and $\mathbf E$ is the diagonal matrix of Ritz energies; the matrices $\mathbf H$ and $\mathbf S$ have elements
\begin{align}
    \textbf{H}_{i,j}&=\bra{\Psi_i}\hat{H}\ket{\Psi_j},\\
    \textbf{S}_{i,j}&=\bra{\Psi_i}\ket{\Psi_j}.
\end{align}
The resulting approximations can systematically approach the exact extremal eigenvalues and eigenfunctions. Preliminary steps to orthogonalize the subspace improve the conditioning of the generalized eigenvalue problem~\cite{epperly2022theory,Zhang2024MeasurementEfficientKrylov,Kirby2024KrylovErrors}.

We restrict the study to K$_1$ and K$_2$, which require moments only through $\hat{H}^3$ and $\hat{H}^5$, respectively. {The first-order Krylov correction constructs the projected Hamiltonian and overlap matrices from $\langle\hat{H}\rangle$, $\langle\hat{H}^2\rangle$, and $\langle\hat{H}^3\rangle$~\cite{Suchsland2021Lanczos,Claudino2021ConnectedMoments}. Sample-based Krylov methods select determinant subspaces from sampled Krylov states~\cite{yu2025quantum,Piccinelli2025RandomizedSQD}; quantum-power and unitary-decomposition approaches provide other Krylov realizations~\cite{Seki2021QuantumPower,Asthana2025QKUD}.}
{Under suitable spectral-separation and reference-overlap conditions, extremal Ritz-value errors can decrease exponentially with Krylov-subspace dimension}~\cite{Saad1980,Kaniel1966,Yoshioka2025,Wagman2025}. In this work, the performance of K$_1$ and K$_2$ is evaluated empirically. {{As we show in Sec.~\ref{sec:results_and_discussion}, applying these corrections after an optimized ADAPT-VQE iteration can reduce the adaptive circuit growth needed to reach a target accuracy on near-term quantum hardware.}}

\subsection{q-sc-EOM}
The expectation-value EOM framework and the self-consistent rotation of operator manifolds originate in classical EOM and propagator theory~{\cite{rowe1968equations,prasad1985some,liu2018unitary}}.
Let $|\Psi(\boldsymbol{\theta})\rangle=U(\boldsymbol{\theta})|\psi_{\mathrm{HF}}\rangle$ denote a variational reference with parameters $\theta$. {q-sc-EOM diagonalizes the Hamiltonian projected into a selected determinant space defined by a set of generators $\hat G_i$~\cite{asthana2023quantum,kumar2023quantum,sundar2026molecular}.} {For the orthonormal generator set used here, the projected eigenvalue problem is}
\begin{equation}
\mathbf K\mathbf c_k=E_k^{(D)}\mathbf c_k.
\end{equation}
Here $\mathbf K$ is the unshifted Hermitian Ritz representation of $\hat H$, with elements
\begin{equation}
K_{ij}=\langle\psi_{\mathrm{HF}}|\hat G_i^\dagger U^\dagger(\boldsymbol{\theta})\hat H U(\boldsymbol{\theta})\hat G_j|\psi_{\mathrm{HF}}\rangle.
\end{equation}
Writing $|\phi_i\rangle=U(\boldsymbol{\theta})\hat G_i|\psi_{\mathrm{HF}}\rangle$, the determinant generators used here give $S_{ij}=\langle\phi_i|\phi_j\rangle=\delta_{ij}$, because the underlying determinants are orthonormal and $U(\boldsymbol{\theta})$ is unitary. Hence $\mathbf S=\mathbf I$ and no sampled overlap matrix is required. A nonorthogonal generator set would instead require the generalized Ritz problem $\mathbf K\mathbf c_k=E_k^{(D)}\mathbf S\mathbf c_k$ after treating linear dependencies in $\mathbf S$~\cite{epperly2022theory,Kwao2026GEP}.

For ground-state correction by subspace diagonalization, we use a $D$-dimensional generator manifold that includes the identity. The projected space then contains the original variational state explicitly and permits it to mix with the selected excited determinants:
\begin{equation}
\hat G_0=\hat I,\qquad \hat{\mathcal G}=\{\hat G_0,\hat G_1,\ldots,\hat G_{D-1}\}.
\end{equation}
The corresponding Ritz states are
\begin{equation}
|\Psi_k^{(D)}(\boldsymbol{\theta})\rangle=\sum_{i=0}^{D-1}c_{ik}\,U(\boldsymbol{\theta})\hat G_i|\psi_{\mathrm{HF}}\rangle.
\end{equation}
With exact matrix elements, let $E_{\mathrm{FCI}}$ denote the full configuration interaction (FCI) ground-state energy of the same active-space Hamiltonian. The Rayleigh--Ritz principle gives
\begin{equation}
E_{\mathrm{FCI}}\leq E_0^{(D)}\leq\langle\Psi(\boldsymbol{\theta})|\hat H|\Psi(\boldsymbol{\theta})\rangle.
\end{equation}
The identity-augmented subspace therefore provides a variational ground-state correction in exact arithmetic: it cannot raise the reference energy or lower it below the active-space FCI energy. Finite-shot matrix estimates can violate these exact-energy bounds even after Hermitization and enforcement of $S=I$. For the lowest eigenvalue of the sampled Hermitian matrix, the sampled reference diagonal element remains an upper bound, but the exact FCI energy need not remain a lower bound.

\begin{figure*}[htbp]
    \centering
    \includegraphics[width=0.9\textwidth]{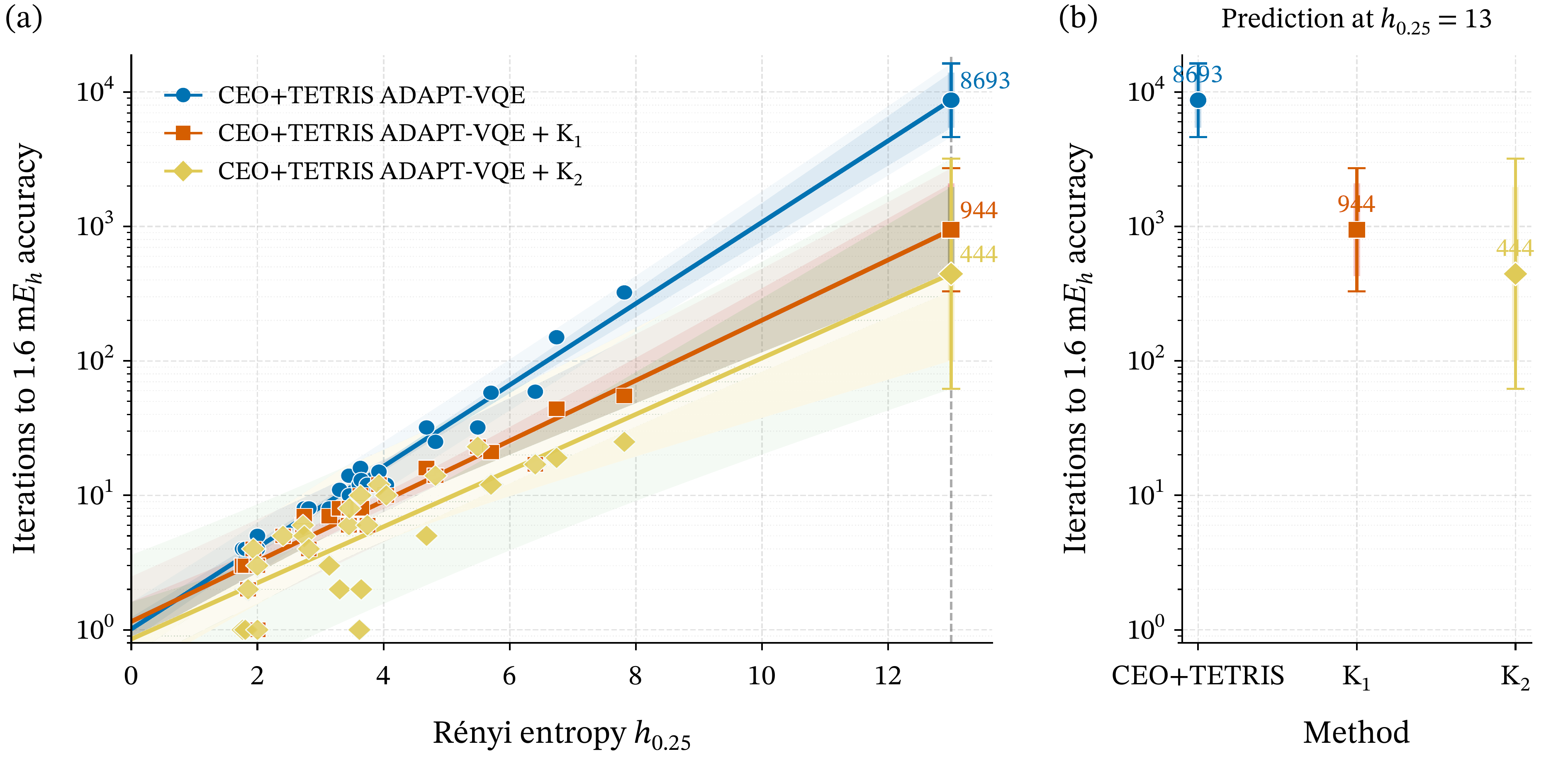}
    \caption{Adaptive outer iterations required to reach $1.6~\mathrm{m}E_h$ accuracy for CEO+TETRIS ADAPT-VQE and its first-order (K$_1$) and second-order (K$_2$) Krylov subspace corrections as a function of R\'{e}nyi entropy $h_{0.25}$. (a) Results for 27 molecular electronic-structure cases. Solid lines are least-squares fits of $\ln N_{\mathrm{iter}}$ to $h_{0.25}$, where $N_{\mathrm{iter}}$ is the iteration count, with $R_{\log}^2=0.970$, $0.865$, and $0.613$ for {CEO+TETRIS ADAPT-VQE}, K$_1$, and K$_2$, respectively. (b) Conditional extrapolations at $h_{0.25}=13$. Darker bands and thick translucent bars denote 95\% confidence intervals for the fitted trend; lighter bands and capped error bars denote 95\% prediction intervals. Observations span $1.764\leq h_{0.25}\leq7.818$; extrapolations assume that the fitted log-linear relation extends beyond this range.}
    \label{fig:krylovadapt}
\end{figure*}

{The enlarged subspace can recover correlation missing from a variational ansatz of limited expressivity using a correction restricted to the reference and single- and double-excitation basis states.} Because adding the identity changes the conventional excited-state q-sc-EOM construction, excitation energies from the truncated identity-augmented space are not guaranteed to retain size intensivity, and the ground-state correction is not generally size extensive across noninteracting fragments. We therefore apply this extension only to the lowest active-space root.

{
\section{Computational details}
\label{sec:computational_details}

All calculations used \textcolor{black}{finite-active-space electronic Hamiltonians and} the publicly available QCANT codebase~\cite{QCANT}. The 27 MolVQE-21~\cite{hagelueken2026exponentialscalingbarriersvariational} cases used Hamiltonians \textcolor{black}{from} Benchmark-QC~\cite{BenchmarkQC}. For systems generated \textcolor{black}{here}, Hartree--Fock references and active-space integrals were obtained with the Python-based Simulations of Chemistry Framework (PySCF)~\cite{pyscf,Pyscf_2020}; fermionic and qubit operators were handled with OpenFermion and PennyLane~\cite{openfermion,bergholm2018pennylane}. \textcolor{black}{The Jordan--Wigner transformation mapped each Hamiltonian to} $N=2n_{\mathrm{orb}}$ untapered qubits, equal to the number of spin orbitals, where $n_{\mathrm{orb}}$ is the number of spatial orbitals.

The MolVQE-21 trajectories \textcolor{black}{used} noiseless state-vector calculations initialized from the Hartree--Fock occupation determinant. CEO-pool ADAPT-VQE \textcolor{black}{employed TETRIS selection~\cite{PhysRevResearch.6.013254,Ramoa2025CEO} and all-parameter reoptimization after each adaptive layer using analytic gradients and the} Broyden--Fletcher--Goldfarb--Shanno (BFGS) algorithm. \textcolor{black}{Reruns used for optimizer-evaluation counts targeted} $\lVert\nabla E\rVert_\infty\leq10^{-8}$, \textcolor{black}{with a 50,000-iteration limit. Exact K$_1$ and K$_2$ energies were obtained by regularized generalized Rayleigh--Ritz diagonalization; implementation details are given in SI Sec.~S5. Accuracy was assessed against FCI for the same active-space Hamiltonian using} $|E-E_{\mathrm{FCI}}|\leq1.6~\mathrm{m}E_h$.

\textcolor{black}{The optimized ADAPT parameters were held fixed during q-sc-EOM calculations. In the hydrogen-chain Davidson--BRG study, both solvers used the same spin-singlet trial space, restricted to states unchanged by reflection through a plane perpendicular to the chain at its midpoint.}

\textcolor{black}{The finite-shot operator-ranking study used} singlet \emph{m}-benzyne in a \textcolor{black}{$(6e,6o)$ active space with the} correlation-consistent polarized valence triple-$\zeta$ (cc-pVTZ) basis. \textcolor{black}{{Standard ADAPT-VQE, without TETRIS selection, appended one operator per iteration from a fermionic singles-and-doubles pool containing 18 singles and 99 doubles.} Each pool gradient was sampled independently using a Gaussian distribution with its exact mean and single-shot variance divided by $N_g$, the number of shots per gradient observable. The operator with the largest sampled magnitude was selected, followed by noiseless BFGS reoptimization with analytic gradients and a gradient tolerance of $10^{-6}$. For each} $N_g\in\{10^3,10^4,10^5,10^6,10^7\}$, \textcolor{black}{100 trajectories were propagated through iteration 44. Energies and the lowest identity-augmented q-sc-EOM root were evaluated noiselessly; the latter used a 118-dimensional trial space in the fixed-$(N_\alpha,N_\beta)$ sector without overlap- or spin-based state tracking. Accuracy attainment was assessed over all evaluated iterations, not used as an online stopping criterion; statistical definitions and sampling details are provided in SI Sec.~S7.}

\textcolor{black}{The separate NH$_3$ finite-shot matrix-sampling study and Davidson--BRG resource model are specified in Appendix~\ref{app:qsceom_resources}. Apart from the two stated sampling studies, calculations excluded sampling and device noise.}
}

\FloatBarrier
\section{Results and discussion}
\label{sec:results_and_discussion}
\subsection{Reduction in ADAPT-VQE iterations using Krylov subspaces}

ADAPT-VQE constructs a problem-adapted ansatz by repeatedly measuring operator-pool gradients, appending selected generators, and reoptimizing the accumulated parameters. Ref.~\cite{hagelueken2026exponentialscalingbarriersvariational} reported an approximately log-linear association between the number of adaptive outer iterations and the R\'{e}nyi entropy for the 21-system MolVQE-21 benchmark. {Here we examine whether applying Hamiltonian-moment Krylov subspace diagonalization starting from the state obtained from running and prematurely stopping CEO-pool TETRIS-ADAPT-VQE can reach $1.6~\mathrm{m}E_h$ accuracy.} We denote by K$_1$ the subspace $\mathrm{span}\{|\psi\rangle,\hat{H}|\psi\rangle\}$ and by K$_2$ the subspace $\mathrm{span}\{|\psi\rangle,\hat{H}|\psi\rangle,\hat{H}^2|\psi\rangle\}$.

Figure~\ref{fig:krylovadapt} shows the first adaptive outer iteration at which $|E-E_{\mathrm{FCI}}|\leq1.6\times10^{-3}\,E_h$ for 27 molecular electronic-structure cases. {The horizontal axis is the determinant-coefficient R\'{e}nyi entropy $h_{0.25}$~\cite{Renyi1961Entropy} in natural logarithmic units, defined in the fixed orbital and occupation-number basis of each Hamiltonian; it is basis-dependent and measures the spread of determinant coefficients.} The benchmark spans $1.764\leq h_{0.25}\leq7.818$, and K$_1$ reduces the iterations needed to reach the threshold by up to approximately sixfold in systems with up to 20 qubits. {Molecular geometries, active spaces, electronic states, and orbital provenance are taken from MolVQE-21~\cite{hagelueken2026exponentialscalingbarriersvariational}, as archived in Benchmark-QC~\cite{BenchmarkQC}.} Least-squares regressions of $\ln N_{\mathrm{iter}}$, where $N_{\mathrm{iter}}$ is the required iteration count, on $h_{0.25}$ give log-space coefficients of determination $R_{\log}^2=0.970$, $0.865$, and $0.613$ for {CEO+TETRIS ADAPT-VQE}, K$_1$, and K$_2$, respectively. {Here, $R_{\log}^{2}$ measures the fraction of variance in $\ln N_{\mathrm{iter}}$ explained by the linear fit.} The fitted slopes are $0.697$ for ADAPT-VQE and $0.516$ for K$_1$, so the descriptive K$_1$ slope is approximately $26\%$ smaller. For H$_{10}$, the highest-entropy case in the sampled range, K$_1$ reduces the required iteration count from 323 to 55, a factor of 5.87 (approximately sixfold). Values reported for $h_{0.25}\geq8$ remain extrapolations beyond the sampled interval. {The entropy and iteration data, regression parameters, and interval definitions for Fig.~\ref{fig:krylovadapt} are provided in Sec.~S5 and \textcolor{black}{Tables~S4--S5} of the Supporting Information.}

{We use the fitted entropy--iteration relation to explore the potential benefits of Krylov corrections for larger correlated systems.} Among the extrapolation targets of Ref.~\cite{hagelueken2026exponentialscalingbarriersvariational}, the [2Fe--2S] cluster in a complete active space of 14 electrons in 12 orbitals [CAS(14e,12o)] and $[\mathrm{Cu}_2\mathrm{O}_2(\mathrm{NH}_3)_6]^{2+}$ in CAS(16e,16o) have $h_{0.25}=10.44$ and $12.67$, respectively, placing the latter essentially at 13. We therefore use $h_{0.25}\simeq13$ as an illustrative extrapolation endpoint for larger active-space problems, including Fe--S clusters and the iron--molybdenum cofactor (FeMoco). {Additional entropy estimates for Fe--S clusters and FeMoco are provided in Sec.~S4 and Table~S3 of the Supporting Information~\cite{LiChan2017SPMPS,Li2019FeMoco,Zhai2026FeMoco}.}

{ At each outer iteration, TETRIS appends a batch of operators with mutually disjoint qubit supports, allowing the operators within that batch to be executed concurrently~\cite{PhysRevResearch.6.013254}. The number of TETRIS batches is therefore a direct measure of the uncompiled adaptive-layer count and provides a qualitative indicator of state-preparation depth. {The compiled batch depth additionally depends on the circuit implementations of operators selected from the CEO pool~\cite{Ramoa2025CEO}, basis changes, gate cancellations, connectivity, and hardware routing.} {Over $h_{0.25}=8$--$13$, extrapolation of the fitted model predicts approximately $3.7$--$9.2$-fold reductions in adaptive iterations (Table~\ref{tab:renyi_iteration_depth_estimates}).}}

\begin{table}[t]\caption{Extrapolated adaptive outer iterations versus R\'{e}nyi entropy. All values at $h_{0.25}=8$--$13$ lie beyond the sampled range $1.764\leq h_{0.25}\leq7.818$ and are exploratory scaling estimates. Both method columns use CEO+TETRIS; errors are 95\% prediction intervals from the 27-case fits.}\label{tab:renyi_iteration_depth_estimates}\tabletext\begin{ruledtabular}\tabletext
\begin{tabular}{ccc}$h_{0.25}$ & ADAPT & K$_1$\\8 & $267^{+162}_{-101}$ & $72^{+86}_{-40}$ \\9 & $536^{+346}_{-210}$ & $120^{+156}_{-68}$ \\10 & $1075^{+746}_{-440}$ & $201^{+284}_{-118}$ \\11 & $2158^{+1614}_{-923}$ & $336^{+521}_{-204}$ \\12 & $4331^{+3504}_{-1937}$ & $563^{+957}_{-354}$ \\13 & $8693^{+7613}_{-4059}$ & $944^{+1763}_{-615}$ \\\end{tabular}\end{ruledtabular}\end{table}

\begin{table*}[t]
\centering
\caption{Finite-range power-law fits for the numbers of nonidentity Pauli terms, $T$, and {qubit-wise-commuting measurement groups}, $G_{h}$, in $\hat{H}$, $\hat{H}^2$, and $\hat{H}^3$. {The fits use $X=C N^{\alpha}$ and a separate set of 25 Hamiltonian cases (19 from MolVQE-21 and six lower-qubit extensions) with $4\leq N\leq12$, where $N$ is the number of qubits.} QWC results are reported for the largest-first (LF), recursive-largest-first (RLF), degree of saturation (DSATUR), and greedy-independent-set (GIS) heuristics. These empirical fits describe scaling within the sampled range.}
\label{Tab:KrylovScalings}
\begin{ruledtabular}
\tabletext
\begin{tabular}{lccccc}
& Raw Pauli terms
& \multicolumn{4}{c}{{QWC measurement groups}} \\
\cline{3-6}
Power & $T$
& $G_{\mathrm{LF}}$
& $G_{\mathrm{RLF}}$
& $G_{\mathrm{DSATUR}}$
& $G_{\mathrm{GIS}}$ \\
\hline
$\hat{H}$
& $0.0677\,N^{3.742}$
& $0.0258\,N^{3.644}$
& $0.0261\,N^{3.637}$
& $0.0267\,N^{3.623}$
& $0.0221\,N^{3.764}$ \\

$\hat{H}^2$
& $9.972{\times}10^{-4}N^{6.944}$
& $9.493{\times}10^{-4}N^{6.082}$
& $9.687{\times}10^{-4}N^{6.067}$
& $1.034{\times}10^{-3}N^{6.028}$
& $7.666{\times}10^{-4}N^{6.290}$ \\

$\hat{H}^3$
& $4.450{\times}10^{-5}N^{8.945}$
& $1.634{\times}10^{-4}N^{7.286}$
& $1.678{\times}10^{-4}N^{7.270}$
& $1.781{\times}10^{-4}N^{7.234}$
& $1.018{\times}10^{-4}N^{7.669}$ \\
\end{tabular}
\end{ruledtabular}
\end{table*}

\subsubsection*{{\textcolor{black}{Quantum computing resource comparison}}}

\begin{figure}
    \includegraphics[width=\linewidth]{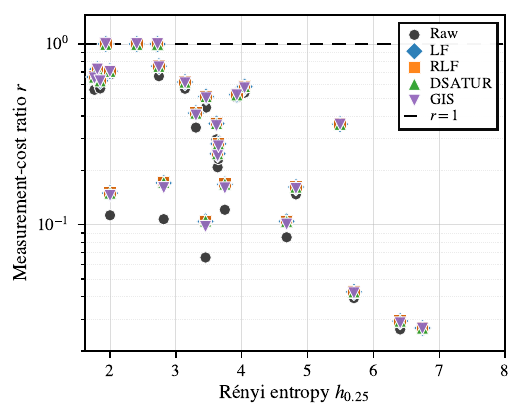}
    \caption{Modeled normalized resource-cost ratio $r=C_{\mathrm{K}_1}/C_{\mathrm{ADAPT}}$ for first-order Krylov subspace diagonalization (K$_1$) relative to full CEO+TETRIS ADAPT-VQE as a function of R\'{e}nyi entropy $h_{0.25}$. The analysis includes 26 cases, and the horizontal dashed line marks $r=1$. K$_1$ is charged once at the retrospectively identified earliest iteration satisfying the $1.6~\mathrm{m}E_h$ threshold; unsuccessful intermediate checks are not charged.}
    \label{fig:ResourceRatio}
\end{figure}

{Figure~\ref{fig:ResourceRatio} shows the modeled normalized resource-cost ratio of K$_1$-assisted CEO-TETRIS-ADAPT-VQE to full CEO-TETRIS-ADAPT-VQE as a function of R\'{e}nyi entropy.} {The Hamiltonian cases, construction of $\hat{H}^2$ and $\hat{H}^3$, qubit-wise-commuting (QWC) grouping procedure, and input counts for the measurement-count fits in Table~\ref{Tab:KrylovScalings} are provided in Sec.~S3 and Table~S2 of the Supporting Information.} Optimizer energy evaluations and Hamiltonian moments are weighted by fitted Pauli-term or {QWC measurement-group counts}, denoted $G_{h}(\hat{H}^p)$ for power $p$~{\cite{Verteletskyi2020MinimumCliqueCover}}. \textcolor{black}{Here, $h\in\{\mathrm{LF},\mathrm{RLF},\mathrm{DSATUR},\mathrm{GIS}\}$ denotes the grouping heuristic: largest-first (LF), recursive-largest-first (RLF), degree of saturation (DSATUR), or greedy-independent-set (GIS).} {The operator-gradient contribution is assigned a normalized cost of $4N^5$ per adaptive iteration, combining the $N^4$ energy-measurement scaling with the approximate equal-precision gradient-shot estimate of Anastasiou and co-workers~\cite{Anastasiou2023Gradients}.}

For adaptive iteration $j$, let $P_j$ be the number of variational parameters. Let $N_A$ be the first adaptive iteration at which ADAPT-VQE without Krylov subspace diagonalization reaches $1.6~\mathrm{m}E_h$ accuracy, and let $\overline N_A$ be the corresponding K$_1$-assisted iteration. For endpoint $e\in\{A,K_1\}$, define $L_A=N_A$ and $L_{K_1}=\overline N_A$. {The cumulative optimizer cost through endpoint $e$, expressed as an equivalent number of energy evaluations, is}
{
\begin{equation}
 C_{\mathrm{opt},e}=\sum_{j=1}^{L_e}\Big[n^{\mathrm{fev}}_{j}+2P_j n^{\mathrm{jev}}_{j}\Big].
\end{equation}
}
\noindent {Here, $n^{\mathrm{fev}}_j$ and $n^{\mathrm{jev}}_j$ denote, respectively, the numbers of objective-function evaluations and complete gradient-vector evaluations recorded during BFGS optimization at iteration $j$.} {Each} complete gradient-vector evaluation is assigned $2P_j$ Hamiltonian expectation-value evaluations. For the real circuits considered here, this convention is compatible with generalized excitation-shift rules~\cite{Kottmann2021FermionicShift}.

With this normalization, the full ADAPT-VQE resource cost is
{
\begin{equation}
 C_{\mathrm{ADAPT}}=4N_A N^5+C_{\mathrm{opt},A}G_{h}(\hat{H}).
\end{equation}
}
\noindent For K$_1$, the projected overlap and Hamiltonian matrices require moments through $\hat{H}^3$. 
Thus, the total normalized resource cost is
{
\begin{equation}
\begin{aligned}
 C_{\mathrm{K}_1}=4\overline N_A N^5+C_{\mathrm{opt},K_1}G_{h}(\hat{H})+G_{h}(\hat{H}^2)+G_{h}(\hat{H}^3).
\end{aligned}
\end{equation}
}
\noindent {In this model, the K$_1$ moment-evaluation cost is charged once at the earliest iteration satisfying the $1.6~\mathrm{m}E_h$ threshold, identified retrospectively from the energy trajectory. {The figure therefore reports a retrospective best-case comparison under the stated resource normalization that assumes prior knowledge of the successful iteration; the assessment is restricted to K$_1$ under this resource normalization.} The normalized operator-gradient cost remains separate from $G_{h}(\hat{H})$ because the cited estimate uses a naive energy-measurement reference.} K$_2$ requires moments through $\hat{H}^5$ and is not included in Fig.~\ref{fig:ResourceRatio}.

The plotted ratio is
{
\begin{equation}
 r=\frac{C_{\mathrm{K}_1}}{C_{\mathrm{ADAPT}}}.
\end{equation}
}
The same grouping method is used in the numerator and denominator.

{Figure~\ref{fig:ResourceRatio} uses exact-state trajectories and the recorded BFGS objective-function and gradient-vector evaluation histories for the 26 benchmark cases with complete optimizer records, excluding H$_{10}$.} The printed fit coefficients in Table~\ref{Tab:KrylovScalings} are used without rounding the resulting $G_{h}(\hat{H}^p)$ values to integers. {The empirical fits in Table~\ref{Tab:KrylovScalings} are used throughout, so their application to systems with more than 12 qubits is an extrapolation.} The modeled resource-cost ratios tend to decrease with qubit count and R\'{e}nyi entropy within this cohort. For every system with $N\geq14$, $r<1$ for all grouping series, and the maximum reduction is approximately fortyfold. The $\hat{H}^2$ and $\hat{H}^3$ moment-evaluation overhead is increasingly offset by avoided later ansatz growth, operator selection, and reoptimization as the adaptive-iteration reduction grows. {The small Krylov diagonalization is done classically.}

{Additional resource-cost data and accounting details are provided in Sec.~S6 and \textcolor{black}{Tables~S6--S7} of the Supporting Information.}

\begin{figure*}[t]
    \centering
    \includegraphics[width=\textwidth]{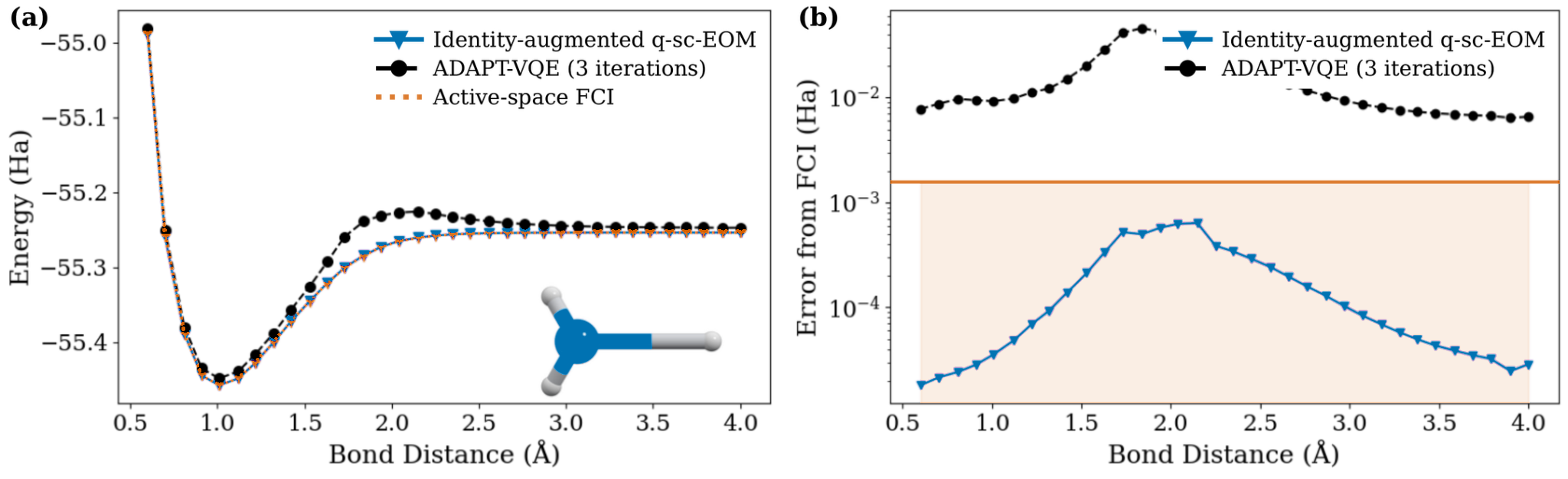}
    \caption{{NH$_3$ ground-state potential energy and absolute error relative to active-space FCI in the STO-6G basis with a $(4e,4o)$ active space, corresponding to eight qubits.} The displayed bond-stretching coordinate spans $0.6$--$4.0~\text{\AA}$. The ADAPT-VQE reference is truncated after three outer iterations, after which q-sc-EOM is applied without reoptimizing the circuit. (a) Energies from truncated ADAPT-VQE, q-sc-EOM, and FCI. (b) Absolute errors $|E-E_{\mathrm{FCI}}|$; the shaded region marks the $1.6~\mathrm{m}E_h$ accuracy threshold. The largest truncated ADAPT-VQE error is approximately $40~\mathrm{m}E_h$, whereas the q-sc-EOM error remains below $1.0~\mathrm{m}E_h$ over the displayed grid. The axis unit Ha denotes the hartree.}
    \label{fig:nh3panel}
\end{figure*}

\subsection{Correction to ground-state energy using identity augmented q-sc-EOM}
{{We next assess identity-augmented q-sc-EOM as a second subspace-diagonalization approach for ground-state energy correction.} The following calculations distinguish energy recovery from truncated references, sensitivity to finite-shot operator ranking, and the measurement and conditioning properties of the projected calculation. In every comparison, q-sc-EOM is applied after the stated number of adaptive outer iterations without reoptimizing the underlying circuit, and FCI denotes exact diagonalization of the same active-space Hamiltonian. We use an absolute energy-error threshold of $1.6~\mathrm{m}E_h$ throughout.

{Figure~\ref{fig:nh3panel} shows the NH$_3$ potential-energy curve in a $(4e,4o)$ active space mapped to eight qubits after three ADAPT-VQE outer iterations.} {Along the sampled bond-stretching coordinate, the truncated ADAPT-VQE error peaks at approximately $40~\mathrm{m}E_h$ near $2.1~\text{\AA}$, whereas the smallest eigenvalue of the Hamiltonian projected onto the q-sc-EOM subspace remains within $1.0~\mathrm{m}E_h$ of active-space FCI over the displayed grid.}}

Additional q-sc-EOM results for the N$_2$ bond-stretching curve are provided in Sec.~S1 and Fig.~S1 of the Supporting Information. {The complementary stretched-H$_4$ convergence, fidelity, spin, and FCI-root analysis is provided in Sec.~S2.}

\begin{figure*}[t]
    \centering
    \includegraphics[width=\textwidth]{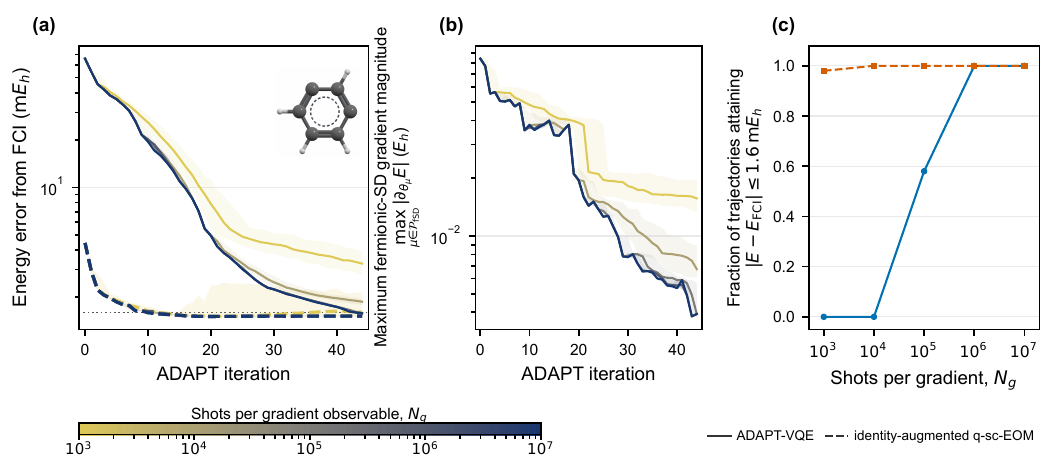}
    \caption{{Effect of finite-shot ADAPT gradient estimation on operator selection for singlet \emph{m}-benzyne mapped to 12 qubits.} (a) Median FCI energy error for ADAPT-VQE (solid) and identity-augmented q-sc-EOM (dashed). (b) Median maximum exact gradient magnitude over the fermionic singles-and-doubles pool. Bands in (a,b) show the 10th--90th percentiles over 100 trajectories, and color denotes shots per gradient observable, \(N_g\). (c) Fraction of trajectories attaining \(\lvert E-E_{\mathrm{FCI}}\rvert\leq 1.6~\mathrm{m}E_h\) at any evaluated ADAPT iteration.}
    \label{fig:mbenzyne_gradient_shots}
\end{figure*}

{Figure~\ref{fig:mbenzyne_gradient_shots} uses finite-shot sampling errors to test operator selection for singlet \emph{m}-benzyne mapped to 12 qubits.} Related ADAPT-VQE analyses identify poor operator selection, operator reordering, and fading operators as mechanisms that can introduce weakly contributing generators~\cite{VaqueroSabater2025PrunedADAPT}. The results show that identity-augmented q-sc-EOM can recover lower energies from ADAPT-VQE states generated by finite-shot operator ranking. They support post-optimization subspace diagonalization when finite-shot-based ranking delays ADAPT-VQE. q-sc-EOM neither estimates nor modifies the ADAPT-VQE pool gradients in this calculation; it obtains a linear variational correction from the state produced by the sampled operator sequence. Each color denotes a fixed number $N_g$ of shots assigned independently to each of the 117 gradient observables in the fermionic singles-and-doubles pool at every ADAPT iteration. Sampling affects which operator is appended; after that selection, parameter reoptimization, the ADAPT-VQE energy, the maximum exact pool-gradient magnitude, and the identity-augmented q-sc-EOM Ritz value are evaluated without sampling noise. {For each shot budget $N_g$, 100 independent ADAPT-VQE trajectories were generated.} Panel~(a) therefore shows the consequences of noisy operator ordering on the median energy error and its 10th--90th percentile range. The solid curves are the ADAPT-VQE energies, and the dashed curves are q-sc-EOM corrections evaluated on the ADAPT-VQE state from the same iteration. Panel~(b) shows how the maximum exact gradient still available in the pool changes along the independently sampled trajectories. The larger late-iteration residual gradients at low $N_g$ coincide with the less accurate trajectories and are consistent with noisy ranking leaving larger descent directions unused at a fixed iteration; the bands do not represent repeated estimates of one fixed gradient.

Panel~(c) summarizes whether each trajectory enters the $1.6~\mathrm{m}E_h$ accuracy window at any evaluated iteration. In ascending $N_g$, the attainment fractions are $(0,0,0.58,1.00,1.00)$ for ADAPT-VQE and $(0.98,1.00,1.00,1.00,1.00)$ for identity-augmented q-sc-EOM. The corresponding fractions within the accuracy window at iteration 44 are $(0,0,0.58,1.00,1.00)$ and $(0.43,0.99,1.00,1.00,1.00)$, respectively. At $N_g=10^3$, 67 q-sc-EOM trajectories leave the window after entering it. Because the projected spaces at successive ADAPT-VQE iterations are not nested, a later q-sc-EOM Ritz value need not preserve an earlier threshold crossing. The complete sampling definition and both attainment summaries are reported in the Supporting Information. Because both methods use identical ADAPT-VQE states generated using noisy operator selection at each iteration, this paired comparison isolates the effect of exact post-optimization q-sc-EOM within the stated hybrid model. Energy evaluation, parameter reoptimization, and q-sc-EOM matrix elements are noiseless, while grouping, covariance, state-preparation cost, and device noise are excluded, defining the scope of the hybrid model in Fig.~\ref{fig:mbenzyne_gradient_shots}.

\subsubsection*{Resource efficiency in identity-augmented q-sc-EOM}

\textcolor{black}{{Davidson finds the target eigenvalue iteratively using Hamiltonian--vector products, while basis-rotation grouping (BRG) groups Hamiltonian terms for measurement in rotated orbital bases. The Davidson--BRG resource-reduction strategy discussed by Sundar et al.\ for excited-state q-sc-EOM~\cite{sundar2026molecular} also applies to the identity-augmented ground-state correction considered here.} Adding the reference determinant increases the trial-space dimension by one without changing the leading-order resource scaling under the same assumptions. Table~\ref{tab:qsceom_resource_scaling} summarizes these scalings, while Appendix~\ref{app:qsceom_resources} provides the ground-state numerical comparisons.}

\begin{table}[t]
\color{black}
\centering
\caption{\textcolor{black}{Formal total measurement-group evaluation scaling for identity-augmented q-sc-EOM, following the resource framework discussed by Sundar et al.~\cite{sundar2026molecular}, the Davidson construction of Kim and Krylov~\cite{kim2023two}, and BRG~\cite{Huggins2021}.}}
\label{tab:qsceom_resource_scaling}
\tabletext
\setlength{\tabcolsep}{4pt}
\begin{ruledtabular}
\begin{tabular}{lc}
Procedure & \shortstack{Measurement-group\\evaluation scaling} \\
\hline
Dense q-sc-EOM, ungrouped & $O(N^{12})$ \\
Dense q-sc-EOM with BRG & $O(N^9)$ \\
Davidson, ungrouped & $O(KN^8)$ \\
Davidson--BRG & $O(KN^5)$ \\
\end{tabular}
\end{ruledtabular}
\end{table}

\textcolor{black}{Here, $N$ is the number of spin orbitals, and $K\geq1$ is the number of Davidson Hamiltonian actions (matrix--vector products) needed to reach the chosen convergence tolerance. These scalings assume that occupied and virtual orbital counts grow proportionally to $N$, and that the retained BRG group count is $O(N)$. The Davidson counts include the initial diagonal-state evaluations; $K$ may depend on $N$. Each prepared state is sampled separately, although the Hamiltonian measurement groups are reusable. Counts describe total measurement-group evaluations over all required prepared states.}

\textcolor{black}{The identity-augmented basis remains orthonormal and therefore retains the absence of overlap-matrix condition-number amplification discussed by Kwao et al.~\cite{Kwao2026GEP}. Hamiltonian sampling errors remain; a ground-state finite-shot comparison with quantum subspace expansion (QSE) is provided in Appendix~\ref{app:qsceom_resources}.}

{
\section{Conclusion}
\label{sec:conclusion}

{We explored the use of first- and second-order Krylov corrections and identity-augmented q-sc-EOM to improve energy accuracy and reduce resource requirements for variational quantum ansätze.} In molecular systems studied with up to 20 qubits, K$_1$ reduces the adaptive iterations needed to reach the $1.6~\mathrm{m}E_h$ threshold by up to approximately sixfold. The benchmark shows a general trend toward larger iteration reductions at higher determinant R\'{e}nyi entropy. {The normalized resource-cost analysis gives reductions of up to approximately fortyfold.} These savings arise from recovering correlation in a compact Krylov subspace and avoiding later rounds of operator selection and parameter reoptimization.

{The NH$_3$ simulations show that identity-augmented q-sc-EOM can recover correlation missing from shallow variational references of limited expressivity without circuit reoptimization.} In the paired \emph{m}-benzyne trajectories, subspace postprocessing attains the $1.6~\mathrm{m}E_h$ accuracy window at earlier ADAPT iterations without altering the finite-shot-selected operator sequence. \textcolor{black}{Identity-augmented q-sc-EOM retains the orthonormal trial basis and permits the use of Davidson--BRG measurement-resource strategies previously discussed for excited-state calculations.} {Both constructions recover correlation through linear Ritz optimization of a fixed variational state. Their subspace sizes, conditioning, and measurement requirements guide the choice of correction. {These constructions could be applied to other unitary variational ansätze beyond ADAPT-VQE, offering routes to improve energy accuracy or reduce the circuit growth and measurement requirements needed to reach a target accuracy. They could also improve the accuracy of properties beyond energy~\cite{Oumarou2025MolecularKrylov}, including molecular response properties calculated using quantum linear response (qLR) theory~\cite{kumar2023quantum}.}}

Future work could develop adaptive strategies for choosing when to apply a subspace correction and how large a subspace to use, guided by the expected energy improvement and measurement cost. Allocating shots according to the sensitivity of the corrected energy to individual moments or matrix elements could further reduce measurement effort. Exploring these strategies in larger active spaces and on quantum hardware could help identify an efficient balance between circuit growth, parameter optimization, and subspace measurements for a target accuracy.
}

{\section*{Supporting Information}}
Additional potential-energy and convergence results; Hamiltonian-power grouping data; determinant-entropy definitions and context; entropy--iteration fits; normalized resource costs and their dependence on accounting assumptions; and operator-ranking statistics (PDF).

\section*{Author Information}
{\subsection*{Notes}}
{The authors declare no competing financial interest.}

{\section*{Acknowledgments}}
This work was supported by National Science Foundation Award No.~2427046. {EB also acknowledges support from the U.S. Department of Energy, Office of Science, Office of Advanced Scientific Computing Research, under Award No. DE-SC0025430.}

{\section*{Data Availability Statement}{\urlstyle{same}The MolVQE-21~\cite{hagelueken2026exponentialscalingbarriersvariational} inputs are available from the Benchmark-QC repository~\cite{BenchmarkQC} at \url{https://github.com/asthanalab/BenchmarkQC}. The QCANT codebase used for the calculations is available at \url{https://github.com/asthanalab/QCANT}.} The Supporting Information provides the tabulated inputs and procedures for the iteration and resource analyses, together with additional convergence and sampling details.}

\appendix
\begingroup
\color{black}
\section{\textcolor{black}{Resource analysis of identity-augmented q-sc-EOM}}
\label{app:qsceom_resources}

\begin{table*}[t]
\color{black}

\centering
\setlength{\tabcolsep}{3pt}
\caption{Measurement-group requirements for projected Davidson q-sc-EOM. \(N\) is the number of qubits, \(D_{\mathrm{full}}\) is the full identity-augmented q-sc-EOM dimension, \(D\) is the dimension of the symmetry-restricted space used for diagonalization, and \(K\) is the number of Davidson Hamiltonian actions. \(G_{\mathrm{QWC}}\) and \(B_\tau\) are the numbers of largest-first QWC and BRG measurement groups per energy evaluation of the same \(\hat{H}_\tau\), respectively. Each \(M\) is the total number of prepared-state/measurement-group evaluations; a group measured on two prepared states contributes two evaluations. If every group receives the same number of shots, ratios of \(M\) also equal total-shot ratios; ratios above one favor Davidson--BRG.}
\label{tab:davidson_brg_resources}
\tabletext
\begin{ruledtabular}
\tabletext
\begin{tabular}{lrrrrrrrrrr}
& & \multicolumn{2}{c}{Trial dimension} & & \multicolumn{2}{c}{\shortstack{Groups per\\energy}} & \multicolumn{2}{c}{\shortstack{Measurement-group\\evaluations}} & \multicolumn{2}{c}{\shortstack{Ratios of measurement-\\group evaluations}} \\
\cline{3-4}\cline{6-7}\cline{8-9}\cline{10-11}
System & \(N\) & \(D_{\mathrm{full}}\) & \(D\) & \(K\) & \(G_{\mathrm{QWC}}\) & \(B_\tau\) & \(M_{\mathrm{dense}}^{\mathrm{BRG}}\) & \(M_{\mathrm{Dav}}^{\mathrm{BRG}}\) & \(\dfrac{M_{\mathrm{dense}}^{\mathrm{BRG}}}{M_{\mathrm{Dav}}^{\mathrm{BRG}}}\) & \(\dfrac{M_{\mathrm{dense}}^{\mathrm{QWC}}}{M_{\mathrm{Dav}}^{\mathrm{BRG}}}\) \\
H$_2$ & 4  & 4   & 2  & 1  & 5   & 4  & 12       & 20       & \(0.600\) & \(0.750\) \\
H$_4$ & 8  & 27  & 9  & 9  & 68  & 8  & 360      & 792      & \(0.455\) & \(3.864\) \\
H$_6$ & 12 & 118 & 30 & 19 & 284 & 15 & \(6\,975\) & \(9\,285\) & \(0.751\) & \(14.223\) \\
H$_8$ & 16 & 361 & 81 & 24 & 851 & 21 & \(69\,741\) & \(43\,029\) & \(1.621\) & \(65.681\) \\
H$_{10}$ & 20 & 876 & 182 & 26 & \(2\,036\) & 28 & \(466\,284\) & \(138\,320\) & \(3.371\) & \(245.124\) \\
\end{tabular}
\end{ruledtabular}
\end{table*}

\paragraph*{Davidson--BRG-based measurement-resource reduction:}

{
{Davidson is an iterative Rayleigh--Ritz eigensolver that expands a trial subspace using residual-based correction vectors~\cite{DAVIDSON197587,qdavidson,reinholdt2024subspace,reinholdt2025pe}. Kim and Krylov developed a Davidson implementation for q-sc-EOM excited-state calculations~\cite{kim2023two}. \textcolor{black}{Davidson algorithms have also been extended for quantum self-consistent linear response (q-sc-LR) calculations of excitation energies and molecular response properties, including polarizable-embedding environments~\cite{reinholdt2024subspace,reinholdt2025pe}.} Basis-rotation grouping (BRG) uses low-rank Hamiltonian factorization and measurements in rotated orbital bases~{\cite{Huggins2021,motta2021low,Cohn_QFD_CDF_2021}}. \textcolor{black}{Sundar et al.\ discussed combined Davidson and BRG as resource-reduction strategies in q-sc-EOM excited-state work~\cite{sundar2026molecular}.} Here we discuss how the ideas of their combined use to reduce measurement resource requirements \textcolor{black}{are} transferable to ground-state corrections in the identity-augmented q-sc-EOM space.}

{
\begin{figure}[t]
    \color{black}
    \centering
    \includegraphics[width=0.45\textwidth]{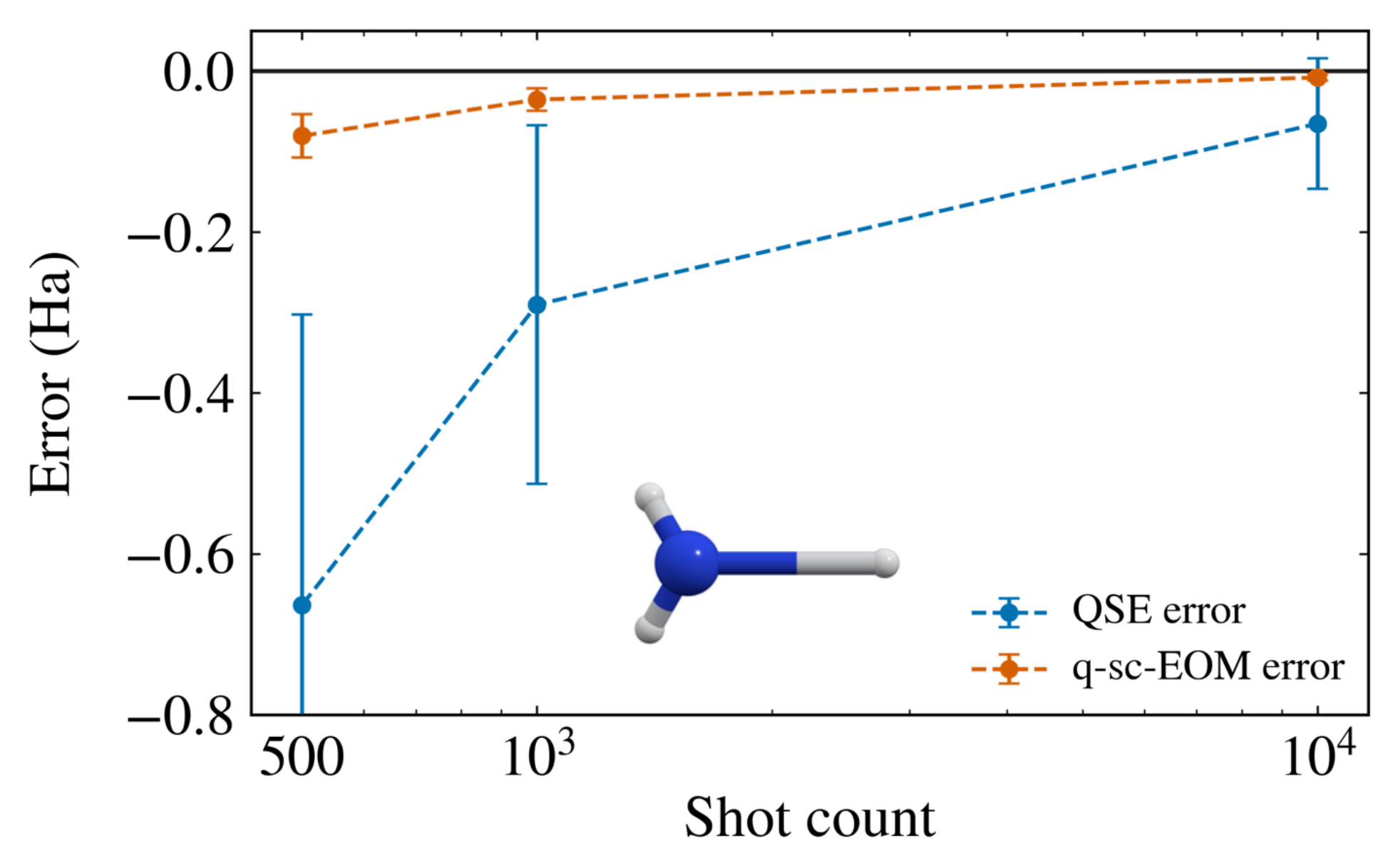}
    \caption{Finite-shot ground-state energy errors for planar NH$_3$ in STO-6G with a $(4e,4o)$ active space. QSE and q-sc-EOM use 500, $10^3$, or $10^4$ shots per measured matrix element. The signed error is $\Delta E=E_{\mathrm{sampled}}-E_{\mathrm{FCI}}$, where $E_{\mathrm{FCI}}$ is the active-space FCI ground-state energy. Symbols show the mean over five independent sampling realizations at each shot count; error bars show $\pm$ one standard deviation. The black horizontal line marks $\Delta E=0$. Only finite-shot sampling is included; gate and readout noise are excluded. The axis unit Ha denotes the hartree.}
    \label{fig:nh3shots}
\end{figure}
}

For an identity-augmented singles-and-doubles q-sc-EOM trial space, \(D=O(n_{\mathrm{orb}}^4)\), so explicit dense construction entails \(D(D+1)/2=O(n_{\mathrm{orb}}^8)\) independent projected-matrix elements, and dense diagonalization requires \(O(D^3)=O(n_{\mathrm{orb}}^{12})\) classical operations. To quantify whether Davidson iteration can reduce the associated quantum measurement burden, we considered neutral linear H$_n$ (\(n=2,4,6,8,10\)) chains at a nearest-neighbor separation of \(1.5~\text{\AA}\) in STO-3G, with all electrons and orbitals active. Every system used the same reference protocol: Hartree--Fock followed by one full-pool fermionic-singles-and-doubles ADAPT selection. The identity-augmented q-sc-EOM trial space contained \(\hat{G}_0=\hat{I}\) together with all spin-conserving single and double excitations. Both solvers used the symmetry-restricted space defined in Computational Details and the same BRG-truncated Hamiltonian \(\hat{H}_\tau\), with \(\tau=10^{-4}\). Here, \(\tau\) denotes the cumulative discarded factor-weight bound used by OpenFermion.

{The lowest root was solved using an unpreconditioned Davidson iteration with residual tolerance $10^{-8}$, at most 200 iterations, a maximum subspace dimension of 128, a restart dimension of 40, and an identity-only initial vector.} {For an unpreconditioned, unrestarted iteration, the residual expansions generate a Krylov subspace of the projected Hamiltonian.} The Davidson iterations evaluated projected matrix--vector products on demand. The dense symmetry-restricted projected matrix was nevertheless constructed to supply diagonal elements for the superposition-state reconstruction of projected Hamiltonian actions and to independently validate the converged root. Let $G_{h}$ denote the number of measurement groups per energy evaluation for grouping method $h$, $B_\tau$ the number of BRG groups at truncation threshold $\tau$, and $K$ the number of Davidson Hamiltonian actions. The dense schedule requires
\begin{equation*}
M_{\mathrm{dense}}^{h}=G_{h}\frac{D(D+1)}{2},
\end{equation*}
whereas the Davidson--BRG model gives
\begin{equation*}
M_{\mathrm{Dav}}^{\mathrm{BRG}}=B_\tau\left[D+K(D+1)\right].
\end{equation*}
{The Davidson count comprises $D$ diagonal-state evaluations and $K(D+1)$ evaluations for the Hamiltonian actions, each measured using $B_\tau$ groups.}
The results define a projected Davidson measurement-resource model validated against dense diagonalization using exact projected matrices.
}

{
{Table~\ref{tab:davidson_brg_resources} quantifies the measurement requirements of this strategy for the identity-augmented ground-state calculations studied here, using the recorded Davidson Hamiltonian-action counts.} Davidson--BRG first reduces the measurement-group evaluation count relative to dense BRG at H$_8$, where the count decreases from \(69\,741\) to \(43\,029\) measurement-group evaluations, a \(38.3\%\) reduction (factor \(1.62\)). At H$_{10}$, the count decreases from \(466\,284\) to \(138\,320\), a \(70.3\%\) reduction (factor \(3.37\)). Relative to dense QWC measurement of the same \(\hat{H}_\tau\), the corresponding reduction factors are \(65.7\) for H$_8$ and \(245.1\) for H$_{10}$; these larger factors combine the benefits of BRG and Davidson and should not be attributed to Davidson alone. All Davidson residual norms are below \(6.1\times10^{-9}\), and BRG truncation changes the lowest root by less than 0.98 microhartree ($0.98~\mu E_h$) within the same q-sc-EOM subspace. The counts exclude {finite-shot estimator variances and covariance-aware shot allocation~\cite{Gonthier2022MeasurementRoadblock,Zhu2024ShotAssignment,vanStraaten2021MetricAwareMeasurement,Zhang2024MeasurementEfficientKrylov,Kirby2024KrylovErrors}, as well as} preparation of general Davidson-vector states, the common correlated-reference circuit, native-gate synthesis, and hardware routing. {In particular, the fixed-reference construction of Ref.~\cite{kim2023two} uses three multiqubit-controlled gates per single excitation and seven per double excitation. A fully supported singles-and-doubles vector then entails \(O(N^4)\) such gates in a sequential construction.} Consequently, the reductions in measurement-group evaluations do not establish corresponding reductions in compiled circuit depth or runtime.

\paragraph*{\textcolor{black}{Avoidance of overlap-matrix condition-number amplification:}}

{\textcolor{black}{QSE has been applied to ground-state energy improvement and error mitigation~{\cite{mcclean2017hybrid,colless2018computation,Yoshioka2022GSE}}. Related virtual-QSE extensions can recover dynamical correlation associated with orbitals outside a restricted active space~\cite{Takeshita2020,motta2024subspace}.} QSE generally solves the generalized problem $\mathbf H\mathbf c=E\mathbf S\mathbf c$. Perturbations in a measured overlap matrix $\mathbf S$ can be amplified when $\mathbf S$ is ill-conditioned, making spectral estimates sensitive to sampling and regularization choices. In q-sc-EOM, orthonormality is analytic and $\mathbf S=\mathbf I$, so $\kappa_2(\mathbf S)=1$, where $\kappa_2$ is the spectral condition number. This removes overlap-matrix condition-number amplification. This has been well studied in Refs.~\cite{Kwao2026GEP,epperly2022theory,Zhang2024MeasurementEfficientKrylov,Kirby2024KrylovErrors}. Here, we show the specific conditioning effect for planar NH$_3$ in the STO-6G basis with a $(4e,4o)$ active space showing that the same improvement of q-sc-EOM also extends to \textcolor{black}{ground-state estimation} using \textcolor{black}{identity-augmented} q-sc-EOM. The Cartesian coordinates used for Fig.~\ref{fig:nh3shots}, in \AA, are N $(0,0,0)$, H $(2.526315789473684,0,0)$, H $(-0.506,0.876,0)$, and H $(-0.506,-0.876,0)$. At the selected geometry, the noiseless QSE overlap matrix has $\kappa_2(\mathbf S)=427.6$. Figure~\ref{fig:nh3shots} reports five independent finite-shot calculations at each of 500, $10^3$, and $10^4$ shots per matrix element; gate and readout noise are not included. The QSE estimates show broader dispersion and stronger shot-count dependence than the q-sc-EOM estimates for this ill-conditioned case. This comparison is consistent with reduced sensitivity when overlap-matrix estimation is avoided, but the different trial spaces and Hamiltonian estimators prevent attribution of the full difference to conditioning alone.}
}

\endgroup

\FloatBarrier
\urlstyle{same}
\bibliography{main}
\end{document}